\documentclass[preprint2]{aastex}

\slugcomment{MSUPHY99.07}

\begin{document}


\title{Determination of meteor showers on other planets using comet 
ephemerides}

\author{Shane L.\ Larson}
\affil{Department of Physics, Montana State University, Bozeman,
Montana 59717}


\begin{abstract}
Meteor showers on the Earth occur at well known times, and are
associated with the decay of comets or other minor bodies whose
orbital paths pass close to the Earth's trajectory.  On the surface,
determining the closest proximity of two orbital paths appears to be a
computationally intensive procedure.  This paper describes a simple
geometric method for determining the proximity between the orbital
paths of two bodies ({\it i.e.}, a comet and a planet) in the solar
system from the known ephemerides of the objects.  The method is used
to determine whether or not meteor showers on other planets in the
solar system could be associated with any of $250$ known comets.
\end{abstract}


\section{INTRODUCTION}\label{sec:intro}
As they traverse their orbits about the Sun, comets slowly evaporate 
and fragment, leaving small bits of cometary debris along their 
orbital tracks.  Some comet orbits intersect the Earth's path, and the 
planet sweeps up a portion of these particulates each year.  
Generally, these particles are drawn into the atmosphere, where they 
burn up at high altitudes, producing the yearly meteor showers.  A 
sample of the meteor showers expected on a regular basis for 
Earth-bound observers is given in Table \ref{EarthShowers}.  A very 
detailed list of meteor streams encountered by the Earth has been 
composed based on ground-based observations of amateur astronomers 
around the world \cite{Jenniskens}.

Given the large number of meteor showers seen on the Earth, it seems 
natural to ask about the possibility of meteor showers on other 
planets.  It may be impractical for a sky-observer of the future to 
view meteor showers from some worlds: Mercury has no atmosphere, the 
clouds of Venus are so thick most meteors will likely burn up before a 
planetbound observer could see them, Jupiter has no solid surface to 
sit on while viewing the shower, and so forth.  Never-the-less, 
predicting regular meteor showers on other worlds may be important for 
protecting explorers and spacecraft from incoming particles, and could 
be useful for planning expeditions and experiments to collect cometary 
material.  It has been suggested \cite{AGM96} that future Mars landers 
may be able to detect meteors from the surface of Mars.

A great deal of modern research has been devoted to analysis of the 
evolution of meteor streams in the solar system, particularly those 
that intersect the Earth's orbit (for example, detailed analyses of 
the evolution of the Quadrantid stream can be found in 
\cite{Hughes79}, \cite{Williams79} and \cite{Murray80}; the Geminid 
stream is analyzed in \cite{Fox82} and \cite{Jones85}).  These 
analyses take into account perturbations to the orbits of the parent 
bodies, as well as the subsequent evolution of the debris trail after 
the comet or minor body has continued on in its orbit.  Over time, 
streams may wander into a planet's path causing new meteor storms, or 
may wander out the planet's path quenching a shower which has been 
periodic for decades or centuries ({\it e.g.}, \cite{Murray80} 
estimates that the Quadrantid shower will vanish by the year 2100).

To a first approximation, however, meteor showers will occur if the
orbit of a planet and the orbit of a minor body intersect (or pass
close to one another).  One way to determine if this occurs is to
evolve the two orbits on a computer and watch for an intersection.  A
generalized method for finding the minimum separation between two
orbits has been described by Murray, Hughes and Williams (1980), but
the method reduces to a coupled set of equations for the orbital
anomalies which requires numerical solution.  Alternatively, the
methods described in this paper approach the problem of determining
the intersection of orbital planes in a completely analytical fashion,
requiring only geometrical methods and matrix algebra.

Section \ref{sec:Orbits} describes the basic parameters and coordinate
systems used to characterize orbits in this paper.  Section
\ref{sec:Rotation} describes the rotations used to correctly orient
two orbits with respect to each other, and applies the rotations to
essential vectors needed for the analysis.  Section
\ref{sec:Intersection} uses the rotated vectors to determine the
intersection between two orbital planes, and computes the distance
between the orbital paths when the planes intersect.  Section
\ref{sec:Shower} proposes a criteria for the existence of a meteor
shower based on the distance between the orbits at intersection.  The
``time'' of showers meeting the criterion is determined.  Section
\ref{sec:Conclusion} applies the condition of Section \ref{sec:Shower}
to 250 known comets, summarizes the results, and discusses the
limitations of determining meteor showers using this method.

Throughout this paper, SI ({\it Syst\`eme Internationale}) units are 
employed, except where the size of the units makes it convenient to 
work in standard units employed in astronomy ({\it e.g.}, on large 
scales, astronomical units (AU) will be used, rather than meters).

\section{Describing Orbits}\label{sec:Orbits}

As is well known, one of the great discoveries of Johannes Kepler was 
that the planets travel on elliptical paths, with the Sun at one focus 
of the ellipse (Kepler's First Law of Planetary Motion, published in 
1609).  Since then, an enormous body of knowledge has been developed 
regarding the analysis of orbital motion (see, for example 
\cite{MarionThornton}), allowing the determination of the position of 
virtually any object in the solar system at any moment in time.

For the work presented here, a time dependent analysis of the orbital
motion is not necessary \footnote{We are interested in knowing the
proximity of two orbits when they cross.  An interesting (but
ultimately more difficult) question to address is whether two bodies
might actually {\it collide} because their orbits intersect.}.  The
only information which is required is a knowledge of the trajectory of
the orbit through space.  The distance of the orbital path from the
Sun may be written for elliptical orbits as
\begin{equation}
    r = {{a(1 - e^{2})}\over{1 + e \cos \theta}}\ ,
    \label{rDistance}
\end{equation}
where $a$ is the semi-major axis of the orbit, $e$ is the 
eccentricity, and $\theta$ is the angle (the anomaly) between the body 
and the axis defined by perihelion, as measured in the orbital plane.  
The perihelion distance for the object can be found from Eq.\ 
(\ref{rDistance}) by taking $\theta = 0$, yielding
\begin{equation}
   r_{p} = a (1 - e) \ .
    \label{Perihelion}
\end{equation}

The distance expressed in Eq.\ (\ref{rDistance}) describes the correct 
size and shape of an elliptical orbit for any object around the Sun, 
but more information is needed to correctly orient the orbit in 
three-dimensional space.  This information is typically collected in 
the orbital ephemeris, which expresses orbital parameters with respect 
to the plane which is coincident with the orbital plane of the Earth 
(the ecliptic).  This paper will use a (cartesian) reference 
coordinate system defined in the ecliptic plane as shown in Figure 
\ref{CoordinateSystem}.  The $+z$ axis is defined perpendicular to the 
ecliptic and in the right handed sense with respect to the Earth's 
orbital motion ({\it i.e.}, when viewed looking down the $+z$ axis, 
the Earth's motion is counter-clockwise in the $xy$-plane).  The $+x$ 
axis is defined along the direction of the Earth's perihelion.

The orbital ephemeris of any body describes its orbit relative to the 
ecliptic plane, and locates the object along its orbital path as a 
function of time.  For the problem of determining the possible 
intersection of two orbital paths, only three elements of the full 
ephemeris for a body will be needed: $\Omega_{o}$ (a {\it modified} 
longitude of the ascending node), $\iota$ (inclination), and $\omega$ 
(argument of perihelion).  Each of these parameters is described 
below, and shown in Figure \ref{AnglesDefine}.

The longitude of the ascending node, $\Omega$, is the angle in the 
ecliptic plane between the vernal equinox (the first point of Aries) 
and the point at which the orbit crosses the ecliptic towards the $+z$ 
direction (``northward'' across the ecliptic).  The parameter, 
$\Omega_{o}$, used in this paper, is an offset longitude measured from 
the perihelion of Earth, rather than the first point of Aries (see 
Figure \ref{OmegaDefine}).  The inclination, $\iota$, is the angle 
between the normal vector of the orbit and the normal vector of the 
ecliptic.  Lastly, the argument of perihelion, $\omega$, is the angle 
between the position of the body as it crosses the ascending node and 
the position at perihelion, as measured {\it in the orbital plane of 
the body}.

In addition to these three angles, it will be useful to define two 
vectors for each orbit of interest: $\hat{n}$, the unit normal vector 
to the plane of the orbit, and $\vec{r_{p}}$, the vector pointing to 
perihelion in the plane of the orbit.

\section{Rotations for Orbital Orientation} 
\label{sec:Rotation}
In order to correctly orient an orbit with respect to the ecliptic, 
assume (initially) that the orbit of interest is in the plane of the 
ecliptic, with the perihelion of the orbit aligned along the 
$+x$ axis ({\it i.e.}, the orbit is co-aligned with the Earth's orbit).  
A series of three rotations, based on the angles $\{\omega, \iota, 
\Omega \}$ from the orbital ephemeris will produce the correct 
orientation.  The first rotation will set the value of the ascending 
node with respect to perihelion, the second rotation will set the 
inclination to the ecliptic, and the third rotation will move the 
ascending node to the correct location in the ecliptic plane.

A useful method for describing rotations is in terms of matrices.  
While it is possible to construct a rotation matrix for rotations 
about a general axis, it is more convenient to conduct rotations about 
the coordinate axes shown in Figure \ref{CoordinateSystem}.  The 
matrices describing rotations about the $x$-, $y$-, and $z$-axes will 
be denoted $\tilde{M}_{x}(\phi)$, $\tilde{M}_{y}(\xi)$, and 
$\tilde{M}_{z}(\psi)$, respectively.

To demonstrate the rotations needed to orient the orbit, consider a 
general vector, $\vec{A}$, which is rigidly attached to the orbital 
plane, maintaining its orientation as the plane is rotated.

The first rotation locates the ascending node with respect to 
perihelion; the rotation depends on the value of the argument of 
perihelion, $\omega$.  This is done by rotating around the $z$-axis by 
$\psi = \omega$.  In terms of rotating a general vector $\vec{A}$, 
this can be written
\begin{equation}
    \vec{A}_{1} = \tilde{M}_{z}(\omega) \vec{A}\ .
    \label{FirstRotation}
\end{equation}
When this operation is applied to the orbit, the ascending node will 
be located on the $+x$ axis.

The orbit is inclined around an axis which passes through the 
ascending node and through the Sun (at one focus of the orbit).  Since 
the first rotation placed the ascending node on the $+x$ axis, and the 
Sun lies at the origin of coordinates, a rotation around the $x$-axis 
by the inclination angle, $\phi = \iota$, will correctly incline the 
orbit.  In terms of the vector $\vec{A}_{1}$ (resulting from Eq.\ 
(\ref{FirstRotation})), this yields
\begin{equation}
    \vec{A}_{2} = \tilde{M}_{x}(\iota) \vec{A}_{1}\ .
    \label{SecondRotation}
\end{equation}

Before the final rotation, it will be convenient to offset the 
longitude of the ascending node such that it is measured from the 
perihelion of the Earth, rather than the vernal equinox (this makes 
the $x$-axis the origin for measuring the longitude of the ascending 
node).  The angle between the vernal equinox and perihelion of Earth 
is simply the argument of perihelion for Earth, $\omega_{\oplus}$, 
giving (see Figure \ref{OmegaDefine})
\begin{equation}
    \Omega_{o} = 2 \pi - \omega_{\oplus} + \Omega\ .
    \label{Offset}
\end{equation}

After the second rotation, the ascending node is still located on the 
$+x$ axis.  Rotation about the $z$-axis by the offset longitude, $\psi 
= \Omega_{o}$, will rotate the longitude of the ascending node to its 
correct location in the ecliptic plane.  In terms of the vector 
$\vec{A}_{2}$(resulting from Eq.\ (\ref{SecondRotation})), this yields
\begin{equation}
    \vec{A}_{3} = \tilde{M}_{z}(\Omega_{o}) \vec{A}_{2}\ .
    \label{ThirdRotation}
\end{equation}
The vector $\vec{A}_{3}$ (which is rigidly attached to the orbit) is 
correctly oriented with respect to the ecliptic.

The two vectors which will be of use later are the unit normal vector to 
the orbit, $\hat{n}$, and the perihelion vector, $\vec{r}_{p}$.  When 
the orbital plane is co-aligned with the Earth's (before any rotations 
have been performed), these vectors have the form
\begin{equation}
    \hat{n} = \left[
        \begin{array}{c}
        0 \\
        0 \\
        1 \\
        \end{array} \right] \qquad, \qquad
    \vec{r}_{p} = \left[
        \begin{array}{c}
        r_{p} \\
        0 \\
        0 \\
        \end{array} \right] \ .
   \label{Vectors}
\end{equation}
The rotation operations described by Eqs.\ (\ref{FirstRotation}), 
(\ref{SecondRotation}), and (\ref{ThirdRotation}) must be applied to these 
vectors so they correctly describe the orbit with respect to the 
ecliptic.  Conducting the rotation procedure yields
\begin{equation}
    \hat{n}' =  \left[
        \begin{array}{c}
        \sin \iota  \sin \Omega_{o}\\
        -\sin \iota \cos \Omega_{o}\\
        \cos \iota \\
        \end{array} \right]\ ,
    \label{OrientedNormal}
\end{equation}
and
\begin{equation}
    {\vec{r}_{p}}\;' = r_{p}\left[
        \begin{array}{c}
        \cos \omega  \cos \Omega_{o} -
        \sin \omega  \cos \iota  \sin \Omega_{o} \\
        
        \cos \omega  \sin \Omega_{o} +
        \sin \omega  \cos \iota \cos \Omega_{o} \\
        
        \sin \omega  \sin \iota \\
        \end{array} \right] \ .
   \label{OrientedPerihelion}
\end{equation}

\section{Intersection of Orbits}\label{sec:Intersection}

The procedure described in Section \ref{sec:Rotation} will correctly 
orient any orbit with respect to the ecliptic.  One could take any 
planet's ephemeris ({\it e.g.}, from the ephemerides given in Table 
\ref{PlanetEphemerides}) and construct the normal vector $\hat{n}$ and 
perihelion vector $\vec{r}_{p}$ in accordance with Eqs.\ 
(\ref{OrientedNormal}) and (\ref{OrientedPerihelion})\footnote{To ease 
the notation, we will drop the primed notation for rotated vectors 
from here on.  It will be understood that the normal vectors and 
perihelion vectors have been correctly oriented with respect to the 
ecliptic.}.  Similar vectors could be generated for cometary 
ephemerides.

The real question of interest is not how the orbital planes of planets 
and comets are related to the ecliptic, but rather how they are 
oriented with respect to each other, and in particular where they 
intersect.  The line defining the intersection of the orbital planes 
can be used to determine whether or not the orbits actually intersect.

Hereafter, assume that vectors related to a comet's orbit will bear the 
subscript `$c$' and vectors related to a planet's orbit will bear the 
subscript `$+$'.  Further, suppose the components of the normal vector 
for a comet's orbit are $\hat{n}_{c} = (a,b,c)$, and the components of 
the normal vector of a planet's orbit are $\hat{n}_{+} = (e,f,g)$.  
Both orbital planes automatically share one point in common: the 
origin, which lies at the focus of each orbital ellipse.  Given this 
point and the two vectors $\hat{n}_{c}$ and $\hat{n}_{+}$, the 
equations describing the two orbital planes are
\begin{equation}
   \begin{array}{lcl}   
             {\rm Comet Plane} & &
	         ax + by + cz = 0\\
             {\rm Planet Plane} & &
	         ex + fy + gz = 0
    \end{array}\ .
    \label{PlaneEquations}
\end{equation}

The intersection of the two planes is a line which is the common 
solution of the two expressions in Eq.\ (\ref{PlaneEquations}).  
Using determinants, the common solution to these equations is found to be
\begin{equation}
    {{x} \over {\left| \begin{array}{cc}   
             b & c \\
	         f & g
          \end{array} \right|}} =
    {{-y} \over {\left| \begin{array}{cc}   
             a & c \\
	         e & g
          \end{array} \right|}} =
    {{z} \over {\left| \begin{array}{cc}   
             a & b \\
	         e & f
          \end{array} \right|}} = k \ ,
    \label{CommonSolution}
\end{equation}
where $k$ is an arbitrary constant.  The solutions $\{x,y,z\}$ of 
Eq.\ (\ref{CommonSolution}) will be points along the line of 
intersection.  It is useful to use these values to define a new 
vector, $\vec{\lambda}$, called the `node vector.'  It points along 
the line of nodes (the intersection of the two planes), and has 
components
\begin{equation}
    \vec{\lambda} = k\left[
        \begin{array}{c}
        bg - cf \\
        ce - ag \\
        af - be \\
        \end{array} \right] \ .
   \label{NodeVector}
\end{equation}

To determine if the orbital paths intersect, one must know the radii 
of the orbits along the line of nodes.  An orbital radius may be 
determined from Eq.\ (\ref{rDistance}) if the value of the anomaly, 
$\theta$, is known.  In terms of two orbits inclined with respect to 
each other, the angles of interest will be the angle between the 
perihelion vector for each orbit, $\vec{r}_{p}$, and the node vector, 
$\vec{\lambda}$.  For each orbit, the angle is defined in terms of the 
dot product of the two vectors, yielding
\begin{equation}
    \cos \theta = {{\vec{r}_{p} \cdot \vec{\lambda}} \over {\left| 
    \vec{r}_{p} \right| \cdot \left| \vec{\lambda} \right|}}\ .
    \label{Anomaly}
\end{equation}
The orbits have two opportunities to intersect: at the ascending node, 
and at the descending node.  Eq.\ (\ref{Anomaly}) gives the angle at a 
single node.  To obtain the value of the anomaly at the other node, 
dot the perihelion vector, $\vec{r}_{p}$, into the negative of 
the node vector, $-\vec{\lambda}$.

Once the anomaly is known, the distance between the orbital paths 
when the planes intersect is simply 
\begin{equation}
    \Delta = | r_{+} - r_{c} | \ ,
    \label{PathSeparation}
\end{equation}
where $r_{+}$ and $r_{c}$ are computed using Eq.\ (\ref{rDistance}) 
with the anomaly defined by Eq.\ (\ref{Anomaly}) and the appropriate 
orbital parameters derived from tabulated ephemerides.

\section{Is there a meteor shower?}\label{sec:Shower}

The occurrence of a meteor shower associated with a particular comet
will depend on the value of the separation between the orbital paths,
$\Delta$.  A variety of proximity criteria could be developed,
depending on specific considerations one would like to make regarding
the likelyhood that particles from a given stream might reach a
planetary body (such as the proximity of the mean orbits of a comet
and planet, or the spread of cometary particles around the orbit of
the parent comet).  The basic criteria may be expressed such that
\begin{equation}
    \Delta \leq \hat{r} \ ,
	\label{TmpCriteria}
\end{equation}
where $\hat{r}$ is the minimum separation one considers likely for a 
shower to occur.

In this paper, the criteria for an orbit intersection causing a meteor
shower will be scaled to a region where the gravitational potential of
the planet dominates over the gravitational potential of the Sun.  The
criteria will be
\begin{equation}
    \Delta \leq \kappa R_{l}\ ,
    \label{criteria}
\end{equation}
where $R_{l}$ is the ``Roche-lobe radius'' (defined as the radius of a
sphere which has the same volume as the planet's Roche lobe) and
$\kappa$ is an arbitrary adjustable scaling factor.  The Roche-lobe
radius can be approximated by
\begin{equation}
    R_{l} \sim 0.52 \cdot a \left[{ m_{+} \over {M_{\odot} + m_{+}} }
    \right]^{0.44} \ ,
    \label{RocheRadius}
\end{equation}
where $m_{+}$ and $M_{\odot}$ are the mass of the planet and the Sun, 
and $a$ is the semi-major axis of the planet's orbit \cite{Iben84}.

Once an intersection (in the sense of Eq.\ (\ref{criteria})) has been
found, one would like to know {\it when} the associated meteor shower
might occur, such that it is possible to mount an observational effort
to detect the shower.  A good marker for a planetary encounter with
the meteor stream is the value of the planetary anomaly along the line
of nodes, given by Eq.\ (\ref{Anomaly}).

Given a particular value of the anomaly, $\theta$, it is possible to 
write a closed-form expression $t(\theta)$ for the time at which the 
planet will arrive at that point in its orbit.  The areal velocity 
may be written
\begin{equation}
   {dA \over dt} = {{\pi a b} \over {\tau}}\ ,
   \label{ArealVelocity}
\end{equation}
where $a$ and $b$ are the semi-major and -minor axes respectively, and
$\tau$ is the period of the orbit.  Separating Eq.\
(\ref{ArealVelocity}) and writing the area element as $dA = (1/2)
r^{2} d\theta$ allows one to write an integral for the time $t$ as
\begin{equation}
   t(\theta) = \left({\tau \over {\pi a b}}\right) {1 \over 2} 
   \int_{0}^{\theta} r^{2} d\theta'\ .
   \label{TimeIntegral1}
\end{equation}
Using the shape equation (Eq.\ (\ref{rDistance})) to write $r = 
r(\theta)$, and using $b = a(1 - e^{2})^{1/2}$ to express the 
semi-minor axis, Eq.\ (\ref{TimeIntegral1}) can be integrated to give
\begin{equation}
   t(\theta) = {\tau \over {2 \pi}}\left[2 \tan^{-1}\left(\sqrt{{1 - 
   e} \over {1 + e}} \tan {\theta \over 2} \right) - {{e \sqrt{1 - 
   e^{2}} \sin \theta} \over {1 + e \cos \theta}}\right]\ ,
   \label{TimeIntegral2}
\end{equation}
which is the time it takes the planet to pass from perihelion
($\theta = 0$) to the anomaly value $\theta$.

\section{Results \& Discussion}\label{sec:Conclusion}

As an example of the methods presented here, a search for comet-planet
orbital intersections in the solar system was carried out using the
planetary ephemerides shown in Table \ref{PlanetEphemerides}
\cite{Standish}, and the comet ephemerides provided in the Jet
Propulsion Laboratory's DASTCOM (Database of ASTeroids and COMets)
\cite{DASTCOM}.  The DASTCOM is a collection of orbital parameters and
physical characteristics for the numbered asteroids, unnumbered
asteroids, and periodic comets, used for analyses of solar system
dynamics.  The DASTCOM includes all known comets with periods less
than $200$ years (periodic comets), and long period comets which have
made a return after $1995$ (there are currently $73$ long period
comets in the DASTCOM, and $208$ comets with periods less than $200$
years).  The orbital elements given in the DASTCOM are osculating 
elements, computed from observations during the comet's most recent 
apparition.

The results of the search for comet-planet orbital intersections are 
listed in Table \ref{Results}, which specified an encounter distance 
of
\begin{equation}
    \Delta \leq 5 R_{l}\ . 
    \label{SearchDelta}
\end{equation}
In all, $128$ possible showers were detected: $3$ at Earth, $1$ at 
Mars, $106$ at Jupiter\footnote{In fact, the results of Table 
\ref{Results} show that Jupiter's orbit intersects the path of comet 
P/Spahr (1998 U4) {\it twice}: one intersection at a separation of 
$\Delta \simeq 1.6 R_{l}$, and a second intersection (at the other 
node) with a separation of $\Delta \simeq 3.3 R_{l}$.}, $17$ at 
Saturn, and $1$ at Uranus.  If one reduces the encounter distance to 
$\Delta \leq 1 R_{l}$, only $32$ possible showers are detected (shown 
at the top of Table \ref{Results}): $1$ at Earth, $28$ at Jupiter, $2$ 
at Saturn, and $1$ at Uranus.  If one allows the encounter distance to 
expand to $\Delta \leq 10 R_{l}$, $188$ possible showers are detected 
(data not shown in Table \ref{Results}): $4$ at Earth, $5$ at Mars, 
$148$ at Jupiter, $24$ at Saturn, $6$ at Uranus, and $1$ at Neptune.

Comets with orbital scales smaller than the solar system (`short
period comets') have evolved largely under the influence of
perturbations due to Jupiter (the mass of Jupiter is greater than the
mass of the other planets combined), giving a large population of
comets which cross Jupiter's orbit.  The search for the origin of
these ``Jovian family comets'' has been a matter of much numerical
simulation and debate (see, for example, \cite{QTD90}).  The
disproportionately large number of showers detected for Jupiter can be
attributed to this feature of the comet population.

A good check of the procedure described in this paper is to consider
the predicted showers at Earth.  In particular, the method outlined in
this work predicts two meteor streams which can be identified with
known showers.  The first is the stream from Comet Tempel-Tuttle,
occurring at $t \sim 318$ d.  This stream can be identified with the
Leonid meteor shower (known to be a stream from Tempel-Tuttle), which
occurs in mid-November each year.  The second is a stream from Comet
Swift-Tuttle, occurring at $t \sim 221$ d.  This stream can be
identified with the Perseid meteor shower (known to be a stream from
Swift-Tuttle), which occurs in mid-August each year.

The possible showers computed here have all assumed that the orbits of 
the comets are static and do not precess.  Further, it is assumed that 
the meteor streams remain attached to those static orbits without 
wandering under the influence of gravitational perturbations in the 
solar system.  In addition, the influences of `local' bodies around 
each planet ({\it e.g.}, Earth's moon, or the Galilean satellites 
around Jupiter) have been ignored.  Never-the-less, the method 
provides a useful way for determining the possibility that a given 
planet will encounter a meteor stream from minor bodies in the solar 
system.

\acknowledgements

I would like to thank M.\ B.\ Larson for comments and suggestions, and
E.\ M.\ Standish who provided helpful discussions regarding planetary
ephemerides.  D.\ K.\ Yeomans and P.\ Chodas provided helpful details
about the DASTCOM database.  I would also like to acknowledge the
hospitality of the Solar System Dynamics Group at the Jet Propulsion
Laboratory during the time this work was completed.  This work was
supported in part by NASA Cooperative Agreement No.\ NCC5-410.

\pagebreak

\begin{figure} 
  \caption{The reference coordinate system in the ecliptic plane.  
  The $z$-axis is defined in the right-handed sense with respect to 
  the Earth's motion, and the $x$-axis points towards the perihelion 
  of the Earth.}
  \label{CoordinateSystem} 
\end{figure}

\begin{figure} 
  \caption{The three essential angles for correctly orienting orbits 
  in three dimensional space are (a) $\Omega_{o}$, the ({\it modified}) 
  longitude of the ascending node; (b) $\iota$, the inclination; and 
  (c) $\omega$, the argument of perihelion.}
  \label{AnglesDefine} 
\end{figure}

\begin{figure} 
  \caption{The {\it modified} longitude of the ascending node, 
  $\Omega_{o}$, defined in terms of the Earth's argument of 
  perihelion, $\omega_{\oplus}$, and the conventional value of 
  $\Omega$ for the orbit.  $\upsilon$ indicates the first point of 
  Aries.}
  \label{OmegaDefine} 
\end{figure}

\begin{figure} 
  \caption{The intersection of two orbits, showing the essential
  quantities for defining the occurrence of a meteor shower: the
  separation of the orbits at crossing, $\Delta$ and the node vector,
  $\vec{\lambda}$, which defines the intersection of the two orbits
  and is used to determine the time of the meteor shower.}
  \label{Intersection} 
\end{figure}


\begin{deluxetable}{ll}
   \tablecaption{Some yearly meteor showers seen from Earth.}
   \tablewidth{0pt}
   \tablehead{
       \colhead{Shower Name} &
       \colhead{Date}
             }
   \startdata
   Quadrantids       & early January  \\
   Lyrids            & mid April      \\
   $\eta$ Aquarids   & early May      \\
   $\delta$ Aquarids & late July      \\
   Perseids          & mid August     \\
   Orionids          & mid October    \\
   Leonids           & mid November   \\
   Geminids          & mid December   \\
 \enddata
 \label{EarthShowers}
\end{deluxetable}

\begin{deluxetable}{lccccc}
   \tablecaption{The mean ephemerides for the planets of the solar system 
                 (epoch J2000).}
   \tablewidth{0pt}
   \tablehead{
       \colhead{Planet} &
       \colhead{$a$} &
       \colhead{$e$} &
       \colhead{$\iota$} &
       \colhead{$\Omega$} &
       \colhead{$\omega$} \\
       \colhead{} &
       \colhead{(AU)} &
       \colhead{} &
       \colhead{$(^{\circ})$} &
       \colhead{$(^{\circ})$} &
        \colhead{$(^{\circ})$}
             }
 \startdata
    Mercury   & 0.38709893 & 0.20563069 & 7.00487 & 48.33167 & 
    29.12478\\
    Venus     & 0.72333199 & 0.00677323 & 3.39471 & 76.68069 & 
    54.85229\\
    Earth     & 1.00000011 & 0.01671022 & 0.00005 & -11.26064 & 
    114.20783\\
    Mars      & 1.52366231 & 0.09341233 & 1.85061 & 49.57854 & 
    286.4623\\
    Jupiter   & 5.20336301 & 0.04839266 & 1.3053 & 100.55615 & 
    -85.8023\\
    Saturn    & 9.53707032 & 0.0541506 & 2.48446 & 113.71504 & 
    -21.2831\\
    Uranus    & 19.19126393 & 0.04716771 & 0.76986 & 74.22988 & 
    96.73436\\
    Neptune   & 30.06896348 & 0.00858587 & 1.76917 & 131.72169 & 
    -86.75034\\
    Pluto     & 39.348168677 & 0.24880766 & 17.14175 & 110.30347 & 
    113.76329\\
 \enddata
 \label{PlanetEphemerides}
\end{deluxetable}

\begin{deluxetable}{llccc}
   \tabletypesize{\footnotesize}
   \tablecaption{The results of a meteor shower search using the comets in 
                 the JPL DASTCOM database.}
   \tablewidth{0pt}
   \tablehead{
       \colhead{Planet} &
       \colhead{Comet} &
       \colhead{($\Delta/R_{l}$)\tablenotemark{a}} &
       \colhead{$\theta$, degrees} &
       \colhead{$t(\theta), d$} 
             }
 \startdata
Earth & 109P/Swift-Tuttle & 0.5043 & 216.49717 & 220.81696\\
Jupiter & P/LONEOS-Tucker (1998 QP54) & 0.02759 & 143.72224 & 1688.24555\\
Jupiter & 117P/Helin-Roman-Alu 1 & 0.05761 & 54.79245 & 605.7507\\
Jupiter & 43P/Wolf-Harrington & 0.10689 & 241.59541 & 2965.94337\\
Jupiter & P/Hergenrother (1998 W2) & 0.13008 & 158.6528 & 1883.37507\\
Jupiter & C/Hale-Bopp (1995 O1) & 0.15907 & 267.71691 & 3287.20798\\
Jupiter & 78P/Gehrels 2 & 0.2102 & 206.19238 & 2510.80676\\
Jupiter & 75P/Kohoutek & 0.21718 & 256.88227 & 3155.62992\\
Jupiter & P/Spahr (1998 W1) & 0.24054 & 267.30471 & 3282.24601\\
Jupiter & 124P/Mrkos & 0.32574 & 344.80757 & 4164.72657\\
Jupiter & 14P/Wolf & 0.32874 & 191.77241 & 2321.02269\\
Jupiter & 53P/Van Biesbroeck & 0.37063 & 143.86166 & 1690.05544\\
Jupiter & 59P/Kearns-Kwee & 0.39771 & 294.48157 & 3602.23635\\
Jupiter & 91P/Russell 3 & 0.41994 & 56.45964 & 624.68681\\
Jupiter & 76P/West-Kohoutek-Ikemura & 0.44039 & 248.71583 & 3054.87529\\
Jupiter & 26P/Grigg-Skjellerup & 0.45251 & 21.59907 & 236.07478\\
Jupiter & 132P/Helin-Roman-Alu 2 & 0.50932 & 176.74286 & 2122.18424\\
Jupiter & P/Kushida (1994 A1) & 0.54051 & 239.05142 & 2933.93263\\
Jupiter & 16P/Brooks 2 & 0.57961 & 175.77613 & 2109.39006\\
Jupiter & 83P/Russell 1 & 0.59462 & 34.84717 & 382.18644\\
Jupiter & D/Kowal-Mrkos (1984 H1) & 0.61747 & 57.35711 & 634.89823\\
Jupiter & 135P/Shoemaker-Levy 8 & 0.68357 & 28.92759 & 316.71554\\
Jupiter & 139P/Vaisala-Oterma & 0.72789 & 241.18336 & 2960.76693\\
Jupiter & 86P/Wild 3 & 0.73742 & 55.46464 & 613.3804\\
Jupiter & 104P/Kowal 2 & 0.81618 & 233.8782 & 2868.46755\\
Jupiter & P/LINEAR-Mueller (1998 S1) & 0.87419 & 157.76011 & 1871.64355\\
Jupiter & P/Shoemaker-Levy 6 (1991 V1) & 0.87826 & 199.12624 & 2418.00496\\
Jupiter & 18P/Perrine-Mrkos & 0.94922 & 228.33725 & 2797.82077\\
Jupiter & 85P/Boethin & 0.97138 & 164.48654 & 1960.19405\\
Saturn & P/Jager (1998 U3) & 0.06673 & 210.03052 & 6366.0197\\
Saturn & 126P/IRAS & 0.53161 & 83.17079 & 2299.91062\\
Uranus & C/Li (1999 E1) & 0.84435 & 137.45475 & 11360.58497\\
\tableline
Earth & 55P/Tempel-Tuttle & 4.15373 & 312.31161 & 318.28395\\
Earth & 26P/Grigg-Skjellerup & 4.518 & 110.36191 & 110.13972\\
Mars & C/LINEAR (1998 U5) & 1.98779 & 90.14579 & 151.61158\\
Jupiter & 47P/Ashbrook-Jackson & 1.00145 & 162.12665 & 1929.08903\\
Jupiter & 15P/Finlay & 1.00753 & 186.90971 & 2256.74179\\
Jupiter & 97P/Metcalf-Brewington & 1.09545 & 175.71325 & 2108.55792\\
Jupiter & 81P/Wild 2 & 1.10168 & 320.59158 & 3897.71867\\
Jupiter & 121P/Shoemaker-Holt 2 & 1.10447 & 264.90153 & 3253.24816\\
Jupiter & 54P/de Vico-Swift & 1.1133 & 152.93248 & 1808.3284\\
Jupiter & 56P/Slaughter-Burnham & 1.1211 & 143.93726 & 1691.03684\\
Jupiter & P/Korlevic-Juric (1999 DN3) & 1.15484 & 347.38897 & 4192.9561\\
Jupiter & P/Mueller 4 (1992 G3) & 1.16212 & 312.27537 & 3804.66421\\
Jupiter & 52P/Harrington-Abell & 1.20057 & 316.89669 & 3856.48162\\
Jupiter & 69P/Taylor & 1.2392 & 274.64237 & 3370.06049\\
Jupiter & P/Larsen (1997 V1) & 1.269 & 224.13556 & 2743.90598\\
Jupiter & 46P/Wirtanen & 1.31801 & 245.23688 & 3011.54806\\
Jupiter & 100P/Hartley 1 & 1.32008 & 21.70502 & 237.23779\\
Jupiter & 87P/Bus & 1.34382 & 15.87435 & 173.3314\\
Jupiter & C/Ferris (1999 K2) & 1.3836 & 285.50894 & 3498.14812\\
Jupiter & 77P/Longmore & 1.43731 & 358.02608 & 4309.06758\\
Jupiter & C/Mueller (1997 J1) & 1.44651 & 262.26797 & 3221.3349\\
Jupiter & P/Hartley-IRAS (1983 V1) & 1.51103 & 166.77415 & 1990.37975\\
Jupiter & 119P/Parker-Hartley & 1.53129 & 236.51823 & 2901.93741\\
Jupiter & 114P/Wiseman-Skiff & 1.54424 & 256.92325 & 3156.13216\\
Jupiter & 102P/Shoemaker 1 & 1.58558 & 143.00259 & 1678.90784\\
Jupiter & P/Spahr (1998 U4) & 1.61195 & 169.07697 & 2020.79437\\
Jupiter & 33P/Daniel & 1.64668 & 229.96736 & 2818.65939\\
Jupiter & 62P/Tsuchinshan 1 & 1.6969 & 261.53064 & 3212.37462\\
Jupiter & 70P/Kojima & 1.78536 & 288.99826 & 3538.80021\\
Jupiter & 60P/Tsuchinshan 2 & 1.84257 & 272.21092 & 3341.08206\\
Jupiter & 4P/Faye & 1.91557 & 192.45922 & 2330.09358\\
Jupiter & 67P/Churyumov-Gerasimenko & 1.95263 & 207.28864 & 2525.16175\\
Jupiter & 6P/d'Arrest & 2.01111 & 126.65145 & 1468.85806\\
Jupiter & 36P/Whipple & 2.02544 & 175.2223 & 2102.06112\\
Jupiter & C/LINEAR (1998 U1) & 2.08822 & 12.83484 & 140.08603\\
Jupiter & C/Spacewatch (1997 BA6) & 2.15772 & 302.67334 & 3696.04414\\
Jupiter & P/Levy (1991 L3) & 2.21811 & 131.97724 & 1536.80051\\
Jupiter & 116P/Wild 4 & 2.27558 & 347.04647 & 4189.21226\\
Jupiter & P/Shoemaker-Levy 1 (1990 V1) & 2.39831 & 215.04783 & 2626.37534\\
Jupiter & 9P/Tempel 1 & 2.41105 & 50.10947 & 552.78419\\
Jupiter & C/LINEAR (1999 H3) & 2.59549 & 138.47987 & 1620.39097\\
Jupiter & 31P/Schwassmann-Wachmann 2 & 2.63332 & 303.86438 & 3709.59249\\
Jupiter & P/LINEAR (1999 J5) & 2.69947 & 98.43266 & 1117.77483\\
Jupiter & C/LINEAR (1998 W3) & 2.71291 & 108.75604 & 1244.38557\\
Jupiter & 21P/Giacobini-Zinner & 2.73222 & 182.72852 & 2201.41502\\
Jupiter & 40P/Vaisala 1 & 2.84407 & 304.29294 & 3714.46212\\
Jupiter & 108P/Ciffreo & 2.88171 & 214.58058 & 2620.30119\\
Jupiter & 136P/Mueller 3 & 2.92721 & 128.57415 & 1493.32993\\
Jupiter & 42P/Neujmin 3 & 2.93144 & 153.24921 & 1812.47528\\
Jupiter & 7P/Pons-Winnecke & 3.13747 & 78.25359 & 876.539\\
Jupiter & 103P/Hartley 2 & 3.19169 & 209.67222 & 2556.32919\\
Jupiter & 128P/Shoemaker-Holt 1-B & 3.22265 & 213.44774 & 2605.56299\\
Jupiter & D/van Houten (1960 S1) & 3.25993 & 336.58556 & 4074.58242\\
Jupiter & P/Kushida-Muramatsu (1993 X1) & 3.26773 & 250.65642 & 3078.93931\\
Jupiter & C/Zhu-Balam (1997 L1) & 3.26781 & 218.82993 & 2675.43314\\
Jupiter & P/LINEAR (1998 VS24) & 3.26871 & 158.78542 & 1885.11843\\
Jupiter & P/Spahr (1998 U4) & 3.29681 & 349.07697 & 4211.40146\\
Jupiter & 61P/Shajn-Schaldach & 3.3833 & 164.21702 & 1956.63978\\
Jupiter & 65P/Gunn & 3.43747 & 49.5381 & 546.34347\\
Jupiter & 120P/Mueller 1 & 3.59615 & 161.60426 & 1922.20873\\
Jupiter & 129P/Shoemaker-Levy 3 & 3.6607 & 284.2452 & 3483.36888\\
Jupiter & 30P/Reinmuth 1 & 3.66617 & 288.52425 & 3533.29093\\
Jupiter & 22P/Kopff & 3.67008 & 113.53434 & 1303.71706\\
Jupiter & 112P/Urata-Niijima & 3.69814 & 194.43253 & 2356.14237\\
Jupiter & 110P/Hartley 3 & 3.70194 & 272.28798 & 3342.00224\\
Jupiter & 49P/Arend-Rigaux & 3.78256 & 288.50492 & 3533.06618\\
Jupiter & P/Lagerkvist (1996 R2) & 3.9786 & 175.43706 & 2104.90302\\
Jupiter & 19P/Borrelly & 4.00335 & 239.67933 & 2941.845\\
Jupiter & 17P/Holmes & 4.00357 & 130.61884 & 1519.42495\\
Jupiter & 137P/Shoemaker-Levy 2 & 4.01952 & 229.51888 & 2812.93064\\
Jupiter & 131P/Mueller 2 & 4.05027 & 208.10376 & 2535.82718\\
Jupiter & P/Jedicke (1995 A1) & 4.07511 & 102.29178 & 1164.85082\\
Jupiter & D/Tritton (1978 C2) & 4.234 & 282.865 & 3467.19321\\
Jupiter & 48P/Johnson & 4.31823 & 104.30891 & 1189.57815\\
Jupiter & 106P/Schuster & 4.41102 & 212.99144 & 2599.62197\\
Jupiter & C/LINEAR (1998 M5) & 4.67076 & 318.4884 & 3874.26616\\
Jupiter & P/LONEOS (1999 RO28) & 4.84285 & 141.12541 & 1654.58458\\
Jupiter & 98P/Takamizawa & 4.88896 & 113.76163 & 1306.55052\\
Jupiter & C/Spacewatch (1997 P2) & 4.89278 & 286.39551 & 3508.49848\\
Jupiter & 73P/Schwassmann-Wachmann 3 & 4.9158 & 51.54895 & 569.03117\\
Jupiter & P/Helin-Lawrence (1993 K2) & 4.99431 & 76.00361 & 850.14018\\
Saturn & C/Catalina (1999 F1) & 1.01329 & 107.67064 & 3035.62842\\
Saturn & P/Gehrels (1997 C1) & 1.27361 & 164.60138 & 4862.59895\\
Saturn & P/Hermann (1999 D1) & 1.35745 & 252.08077 & 7703.6729\\
Saturn & C/LINEAR (1998 Q1) & 1.36923 & 70.34971 & 1928.09742\\
Saturn & P/Shoemaker 4 (1994 J3) & 2.45512 & 358.40089 & 10704.16966\\
Saturn & P/Montani (1997 G1) & 2.89611 & 185.28975 & 5549.17747\\
Saturn & P/Helin (1987 Q3) & 3.16833 & 76.81205 & 2114.42428\\
Saturn & 63P/Wild 1 & 3.2706 & 260.23561 & 7952.44828\\
Saturn & 140P/Bowell-Skiff & 3.30576 & 231.84877 & 7070.65503\\
Saturn & D/Bradfield 1 (1984 A1) & 3.57748 & 262.73062 & 8027.80737\\
Saturn & 134P/Kowal-Vavrova & 3.64885 & 319.9063 & 9665.71078\\
Saturn & C/Spacewatch (1997 BA6) & 3.95068 & 44.94144 & 1214.46702\\
Saturn & C/LINEAR (1999 N4) & 3.95944 & 78.38428 & 2160.08505\\
Saturn & C/LINEAR (1999 H3) & 4.19679 & 241.0887 & 7362.46033\\
Saturn & P/Lagerkvist-Carsenty (1997 T3) & 4.90066 & 120.32592 & 3428.86878\\

 \tablenotetext{a}{The first $32$ entries (above the line) 
                are for intersections having $\Delta < R_{l}$.  All other 
                encounters are for $\Delta < 5 R_{l}$.}
 \enddata
 \label{Results}
\end{deluxetable}

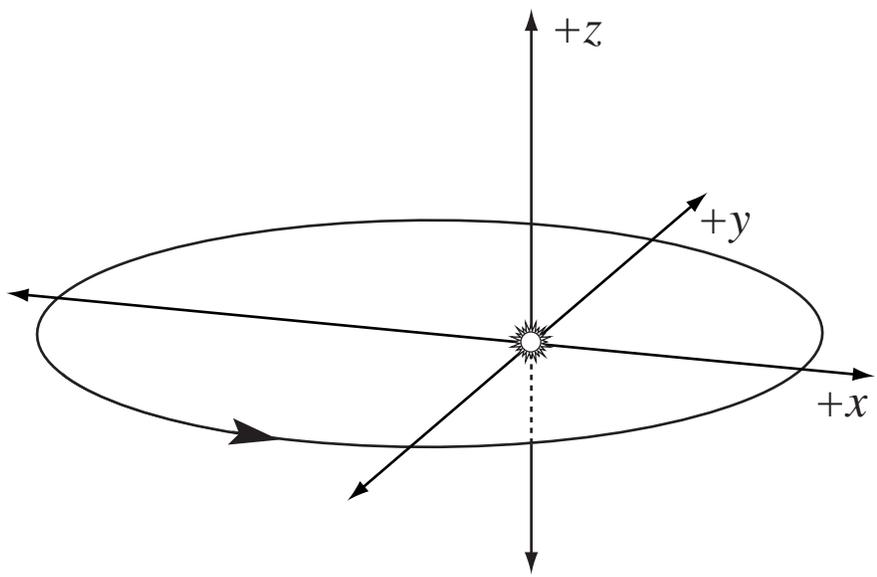
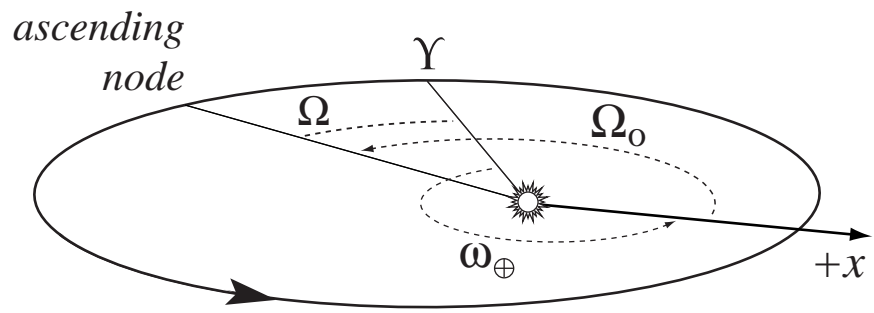
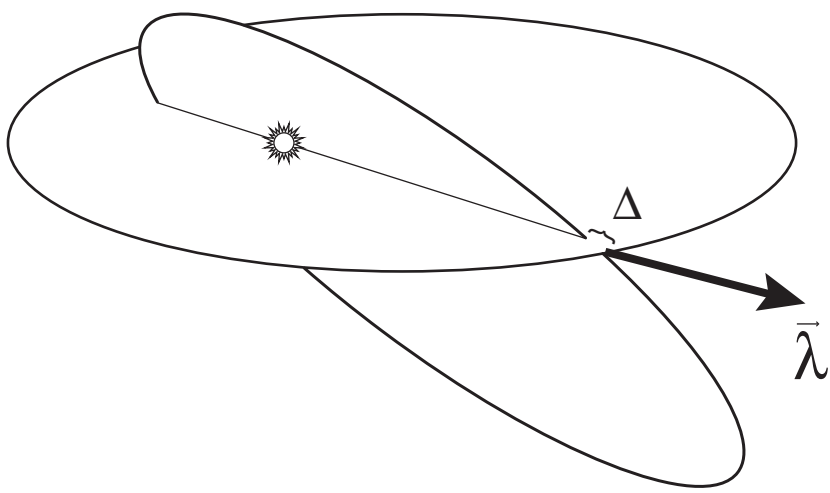
\end{document}